\documentclass[letterpaper, 10 pt, conference]{ieeeconf}  
\usepackage{xcolor}
\usepackage{siunitx}
\usepackage{booktabs}
\usepackage{mathtools}
\usepackage{amssymb}
\usepackage{cite}
\usepackage{graphicx}
\usepackage{epstopdf}
\usepackage{subfig}
\usepackage{multirow}

\IEEEoverridecommandlockouts                              

\title{\LARGE \bf
Equivalent-Circuit Thermal Model for Batteries with One-Shot Parameter Identification
}

\author{Myisha A. Chowdhury$^{1}$ and Qiugang Lu$^{1,\dagger}$
\thanks{*This work was supported by the Texas Tech University and NSF Grant 2340194.}
\thanks{$^{1}$M.A. Chowdhury and Q. Lu are with the Department of Chemical Engineering, Texas Tech University, Lubbock, TX 79409, USA.   
Email: {\tt\small myisha.chowdhury@ttu.edu; jay.lu@ttu.edu}}%
\thanks{$^{\dagger}$Corresponding author: Q. Lu.}%
}

\begin{document}

\maketitle
\thispagestyle{empty}
\pagestyle{empty}

\begin{abstract}
Accurate state of temperature (SOT) estimation for batteries is crucial for regulating their temperature within a desired range to ensure safe operation and optimal performance. The existing measurement-based methods often generate noisy signals and cannot scale up for large-scale battery packs. The electrochemical model-based methods, on the contrary, offer high accuracy but are computationally expensive. To tackle these issues, inspired by the  equivalent-circuit voltage model for batteries, this paper presents a novel equivalent-circuit electro-thermal model (ECTM) for modeling battery surface temperature. By approximating the complex heat generation inside batteries with data-driven nonlinear (polynomial) functions of key measurable parameters such as state-of-charge (SOC), current, and terminal voltage, our ECTM is simplified into a linear form that admits rapid solutions. Such simplified ECTM can be readily identified with one single (one-shot) cycle data. The proposed model is extensively validated with benchmark NASA, MIT, and Oxford battery datasets. Simulation results verify the accuracy of the model, despite being identified with one-shot cycle data, in predicting battery temperatures robustly under different battery degradation status and ambient conditions. 
\end{abstract}

\section{Introduction}
\label{sec:introduction}
Battery-based energy storage systems have become a popular choice in applications like electric vehicles and portable devices, owing to the high energy density, power density, long lifecycle, and low self-discharge rate of modern (e.g., Lithium-ion) batteries \cite{wu2022research}. This widespread usage has mandated the development of advanced battery management systems to ensure safety and reliability. In particular, a thermal management system (TMS) is essential for accurately estimating and regulating battery temperatures for its optimal performance. For example, as the temperature rises, battery performance can initially improve due to enhanced reaction kinetics; however, the prolonged exposure to elevated temperatures can lead to faster degradation and safety risks, including thermal runaway. Conversely, operating at lower temperatures increases the internal resistance and reduces the capacity \cite{richardson2015sensorless}. Moreover, batteries encounter diverse operating conditions in real-world applications, such as rapid charging and discharging cycles, high current loads, and extreme ambient temperatures. These factors directly affect their thermal behaviors. Thus, accurately estimating the state of temperature (SOT) is critical for the battery's health. Traditionally, sensors are employed to measure battery temperatures directly, but the incurred high costs for large-scale battery packs and concerns about reliability and accuracy make them a less appealing solution \cite{naguib2021accurate}. Recently, SOT estimation methods, such as electrochemical impedance spectroscopy (EIS)-based methods \cite{beelen2015improved} and model-based methods, have been explored. For the EIS-based method, AC currents are introduced into the battery, and the temperature is estimated based on the voltage output. Although this method mitigates the limitations of sensor-based methods, it is challenging to adopt it for online SOT estimation due to the burden on the hardware design \cite{naguib2021accurate}. An alternative to direct measurement-based techniques is the model-based methods. 

Model-based methods can be further classified into electrochemical or physics-based models and electro-thermal models. Specifically, electrochemical models aim to simulate temperature profiles under different operating conditions by capturing internal dynamics at the microscopic level. While these models offer high accuracy, they are unsuitable for real-time TMS due to their computational complexity \cite{kumar2022model}. Moreover, they often require detailed measurements of various material properties \cite{samanta2021comprehensive}. To address these shortcomings, researchers have adopted equivalent-circuit electro-thermal models (ECTMs), which simultaneously consider the electrical characteristics and thermal dynamics of the battery, enabling the accurate temperature prediction in real-time \cite{shen2024accurate, liu2023electric, kumar2022model, mahboubi2022developing}. The electrical component is useful for representing the intricate battery dynamics with interconnected components like resistors and capacitors, which can be augmented to capture the nonlinear behaviors. Meanwhile, the thermal model provides insight into the generation and transfer of heat in the battery.

One major issue for using ECTM to describe battery thermal behaviors is the determination of the heat generation during battery charging-discharging cycles \cite{wang2019calculation}. Earlier works determine the heat generation in batteries mainly by performing experimental tests, which are expensive \cite{vertiz2014thermal}. Recently, different models have been developed to estimate the generated heat \cite{mahboubi2022developing, kumar2022model}. In \cite{kumar2022model}, an ECTM is proposed alongside a heat generation model. However, the model exhibits bias in temperature estimation due to its focus on irreversible heat generation, neglecting the contributions of reversible heat. The work in \cite{mahboubi2022developing} presents a second-order ECTM model including both reversible and irreversible heat generation, with findings showing that the heat generation rate and thermal properties are influenced by ambient temperature, current, and state-of-charge (SOC). However, these works rely on detailed analytical modeling of the heat generation that requires the knowledge of material properties, which is challenging to acquire in real-life applications \cite{tran2019computationally}. 

In this work, we develop a new and effective ECTM that can accurately predict battery surface temperatures across a wide range of ambient conditions. Our major contribution lies in the proposition of a physics-based, data-driven heat generation model to describe battery thermal behaviors without requiring direct measurements of the material and thermal properties of batteries. Further, the parameter estimation of the model is formulated into a least-squares problem that can be readily solved with \textit{one-shot} cycling data and is simple enough for real-time application. Finally, the proposed ECTM model is rigorously validated with three benchmark battery datasets under various operating conditions.

The rest of the paper is organized as follows. Section \ref{sec:modeling} introduces the equivalent-circuit model (ECM) and proposes our new ECTM scheme for modeling battery surface temperatures. Section \ref{sec:parameter_identification} presents the least-squares method adopted for identifying model parameters. In Section \ref{sec:results}, we test the effectiveness of the proposed ECTM with three different benchmark battery datasets.

\section{Battery Equivalent-Circuit Modeling}
\label{sec:modeling}
In this section, we present a comprehensive model that integrates both electrical and thermal components alongside a detailed representation of the heat generated within the battery during charging and discharging cycles.

\subsection{ECM for Modeling Battery Voltage}
\label{subsec:ecm}
\begin{figure}[tbh]
\centering
\includegraphics[width=0.9\columnwidth]{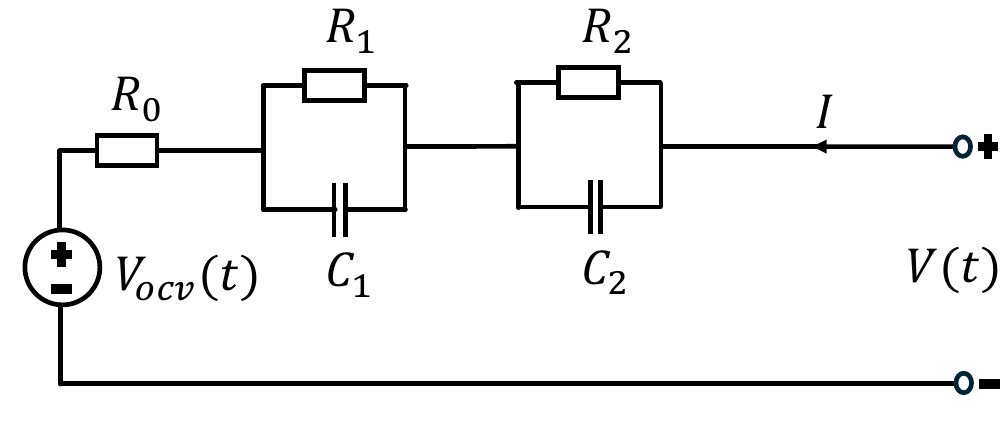}
\caption{The standard Thevenin equivalent-circuit model.} \label{fig:ecm}
\end{figure}
The ECM is a fundamental and simple method for modeling and approximating a battery's terminal and open-circuit voltage (OCV). Fig. \ref{fig:ecm} illustrates the commonly used second-order Thevenin equivalent-circuit model \cite{tian2020one}. One principal component of the model is the OCV: $V_{ocv}(t)$, which represents the maximum voltage that a battery can supply under zero load. The OCV is majorly affected by battery's SOC, state-of-health (SOH), and temperature $T(t)$, i.e., $V_{ocv}(t) = f(SOC(t), SOH(t), T(t))$. The SOC of a battery can be estimated using the definition
\begin{equation}
SOC(t) = SOC(t_0)+\frac{\int_{t_0}^{t}i(t)dt}{Q_{0}}, \label{eq:soc}
\end{equation}
where $t$ is the time, $Q_0$ is the maximum available capacity, and $i(t)$ is the battery current (negative for discharging and positive for charging). The SOC-OCV correlation can be determined experimentally by charging and discharging the battery under a small or normal magnitude of current at large intervals \cite{tran2021comprehensive}. However, the SOC-OCV experiment is time-consuming and expensive to conduct \cite{tian2020one}. Therefore, $V_{ocv}(t)$ is often approximated numerically by a nonlinear (e.g., polynomial) function of the SOC \cite{tian2020one,weng2014unified} as
\begin{equation}
V_{ocv}(t) = \sum_{j=0}^{n}\alpha_j SOC^j(t), \label{eq:ocv}
\end{equation}
where $\alpha_j$, $j = \{0,\ldots,n\}$, are the coefficients. In Fig. \ref{fig:ecm}, $R_0$ represents the ohmic resistance. $R_1$, $R_2$, and $C_1$, $C_2$ in the figure stand for the polarization resistance and capacitance, respectively. 


\subsection{ECTM for Modeling Battery Temperature}
\label{subsec:thermal_model}
\begin{figure}[tbh]
\centering
\includegraphics[width=0.9\columnwidth]{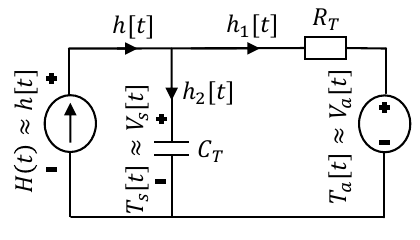}
\caption{Schematic of the ECTM for battery.} \label{fig:thermal}
\end{figure}
Fig. \ref{fig:thermal} shows the presented first-order ECTM that will be used to model the surface temperatures of batteries (with a similar idea as \cite{kumar2022model}). The heat generated at time $t$ is denoted as $H(t)$ and is analogous to the current flowing through the circuit in Fig. \ref{fig:ecm}. Battery surface temperature $T_s(t)$ and ambient temperature $T_a(t)$ are analogous to the voltage across the $C_T$ and $R_T$, respectively. Note that $T_a(t)$ acts as the voltage sink, and $R_T$ and $C_T$ are connected in parallel. The current or heat flowing through the capacitor $C_T$ is denoted as $h_2(t)$ and can be expressed as:
\begin{equation}
h_2(t) = h(t)-h_1(t), \label{eq:current}
\end{equation}
where $h_1(t)$ is the current through the resistor. Replacing $h_2(t)$ and $h_1(t)$ in \eqref{eq:current} by $C_T \dfrac{dT_s(t)}{dt}$ and $\frac{T_s(t)-T_a(t)}{R_T}$ gives
\begin{equation}
C_T \dfrac{dT_s(t)}{dt} = h(t)-\frac{T_s(t)-T_a(t)}{R_T}.
\end{equation}
Rearranging the above equation, we can obtain
\begin{equation}
\dfrac{dT_s(t)}{dt} = -\frac{1}{R_TC_T}T_s(t)+\frac{1}{R_TC_T}T_a(t)+\frac{1}{C_T}h(t). \label{eq:model_diff}
\end{equation}
By discretizing \eqref{eq:model_diff} with sampling interval $\Delta t$, one has
\begin{equation}
	T_s[k] = \epsilon T_s[k-1] + (1-\epsilon) T_a[k-1] + (1-\epsilon) R_T h[k-1] \label{eq:model_disc},
\end{equation}
where $k$ represents the discrete timestep, and $\epsilon := e^{\frac{-\Delta t}{R_TC_T}}$. Note that the heat $h[k]$ at timestep $k$ is generated by electrochemical reactions within the battery during its charging and discharging cycles. The generated heat consists of two sources: reversible heat $h^{re}[k]$ and irreversible heat $h^{ir}[k]$, and can be expressed as \cite{shen2024accurate, mahboubi2022developing} (for notation simplicity, we have omitted timestep $k$): 
\begin{align}
h = h^{ir}+h^{re},h^{ir}:= I(V-V_{ocv}), h^{re} := IT_s\frac{\partial V_{ocv}}{\partial T_s},  \label{eq:heat} 
\end{align}
where $\frac{\partial V_{ocv}}{\partial T_s}$, $V$, and $I$ are entropic coefficient, terminal voltage, and current, respectively. Therefore, eqn. \eqref{eq:heat} can be rearranged into
\begin{equation}
h = I \left[V-\left(V_{ocv}-T_s\frac{\partial V_{ocv}}{\partial T_s}\right)\right]. \label{eq:heat_rearr}
\end{equation}
Note that $V_{ocv}$ and $T_s\frac{\partial V_{ocv}}{\partial T_s}$ above are typically not accessible. We propose to approximate $V_{ocv}$ in \eqref{eq:heat_rearr} by a polynomial function of SOC as \eqref{eq:ocv}. Moreover, as verified in \cite{mahboubi2022developing}, the term $\frac{\partial V_{ocv}}{\partial T_s}$ can also be approximated similarly with a polynomial function of SOC, where at time step $k$, $T_s[k-1]$ is constant:
\begin{equation}
T_s\frac{\partial V_{ocv}}{\partial T_s} = \sum_{j=0}^{n}\beta_j SOC^j. \label{eq:entropic_coefficient}
\end{equation}
Our \textit{unique contribution} in this work lies in the creative integration of the simplified equivalent-circuit structure in \cite{kumar2022model} and the polynomial approximation of heat generation with SOC as in \cite{mahboubi2022developing}. Compared with the two-state thermal model in \cite{mahboubi2022developing}, we propose a simplified, uniform lumped thermal model whose parameters can be identified with one-shot cycle data without requiring any prior knowledge of material properties, such as the thermal capacity. Further, we approximate $V_{ocv}$ as a nonlinear function of SOC as in \eqref{eq:ocv} rather than a linear function of battery core temperature (not available) as done in \cite{mahboubi2022developing}. Here, for simplicity, we choose the same order for polynomials in \eqref{eq:entropic_coefficient} and \eqref{eq:ocv}. Combining \eqref{eq:ocv} and \eqref{eq:entropic_coefficient}, and plugging into \eqref{eq:heat_rearr} give
\begin{equation}
h = I \left[V-\sum_{j=0}^{n} \eta_j SOC^j\right], \label{eq:heat_gen}
\end{equation}
where $\eta_j = \alpha_j-\beta_j$, $j={0,\ldots,n}$, is the coefficient.

Finally, we plug the expression of $h[k-1]$ as in \eqref{eq:heat_gen} into \eqref{eq:model_disc}. After algebraic manipulations, one obtains
\begin{align}
&T_s[k] = \epsilon T_s[k-1] + (1-\epsilon) T_a[k-1] + (1-\epsilon) R_T I[k-1]  \nonumber \\
&~~~ V[k-1] -I[k-1]\sum_{j=0}^{n} (1-\epsilon) R_T\eta_j SOC^j[k-1] \label{eq:model_disc_heat}. 
\end{align}
The involved model parameters to be determined from experimental data consist of 
\begin{equation}
\Gamma = [R_T \quad C_T \quad \eta_0 \quad \eta_1 \quad \eta_2 \quad \eta_3 \quad \ldots \quad \eta_n]^\top \nonumber.
\end{equation}
Note that the model in \eqref{eq:model_disc_heat} is highly nonlinear in $\Gamma$, which may pose significant challenges for parameter identification, especially for real-time applications \cite{zhang2021taking}. To address the issue, we redefine a set of new parameters $\theta_i$ that combines the original parameters as in Table \ref{table:mapping}, which shows an example of $i=\{1,\ldots,9\}$, i.e., $n=6$. This is under an assumption that the sample interval $\Delta t$ in $\epsilon$ is {approximately constant}, as often the case in practical sampling devices. With new parameters defined in Table \ref{table:mapping}, the temperature model in \eqref{eq:model_disc_heat} can be re-written as
\begin{align}
T_s[k] =& \theta_1 T_s[k-1] + \theta_2 T_a[k-1] + \theta_3 I[k-1] V[k-1]-   \nonumber \\
& I[k-1] \left(\sum_{i=1}^{n}\theta_{3+i}SOC^{n-1}[k-1]\right) \label{eq:linear_model}.
\end{align}
Taking $n=6$ as an example, the new parameter set
\begin{equation}
\Theta := [\theta_1 \quad \theta_2 \quad \theta_3 \quad \theta_4 \quad \theta_5 \quad \theta_6 \quad \theta_7 \quad \theta_8 \quad \theta_9]^\top \nonumber.
\end{equation}
Note that with above parameter transformation, the temperature model \eqref{eq:linear_model} becomes linear in terms of $\Theta$, which significantly simplifies the corresponding parameter identification problem. With the defined coefficients $x_{j}[k-1]$ in Table \ref{table:mapping}, the above model in \eqref{eq:linear_model} can be compactly written as 
\begin{equation}
T_{s}[k]=\sum_{j=1}^{m}\theta_{j}x_{j}[k-1]. \label{eq:linear_compact} \end{equation}
where $m=n+3$ is the total number of parameters.

\begin{table}[tbh]
\centering
\caption{Redefinition of the parameters in ECTM \eqref{eq:model_disc_heat}.}
\label{table:mapping}
\begin{tabular}{lll}
\hline
Original & Redefined & \multirow{2}{*}{Coefficients} \\
Parameters & Parameters & \\
\hline
$\epsilon$ & $\theta_1$ & $x_{1}[k-1] \coloneqq T_s[k-1]$\\
$1-\epsilon$ & $\theta_2$ & $x_{2}[k-1] \coloneqq T_a[k-1]$\\
$(1-\epsilon)R_T$ & $\theta_3$ & $x_{3}[k-1] \coloneqq I[k-1] V[k-1]$\\
$(1-\epsilon)R_T \eta_0$ & $\theta_4$ &  $x_{4}[k-1] \coloneqq I[k-1]$\\
$(1-\epsilon)R_T \eta_1$ & $\theta_5$ & $x_{5}[k-1] \coloneqq I[k-1]SOC[k-1]$\\
$(1-\epsilon)R_T \eta_2$ & $\theta_6$& $x_{6}[k-1] \coloneqq I[k-1]SOC^2[k-1]$\\
$(1-\epsilon)R_T \eta_3$ & $\theta_7$& $x_{7}[k-1] \coloneqq I[k-1]SOC^3[k-1]$\\
$(1-\epsilon)R_T \eta_4$ & $\theta_8$& $x_{8}[k-1] \coloneqq I[k-1]SOC^4[k-1]$\\
$(1-\epsilon)R_T \eta_5$ & $\theta_9$& $x_{9}[k-1] \coloneqq I[k-1]SOC^5[k-1]$\\
\hline
\end{tabular}%
\end{table}

\begin{figure*}[tbh]
\centering
\includegraphics[width=1.9\columnwidth]{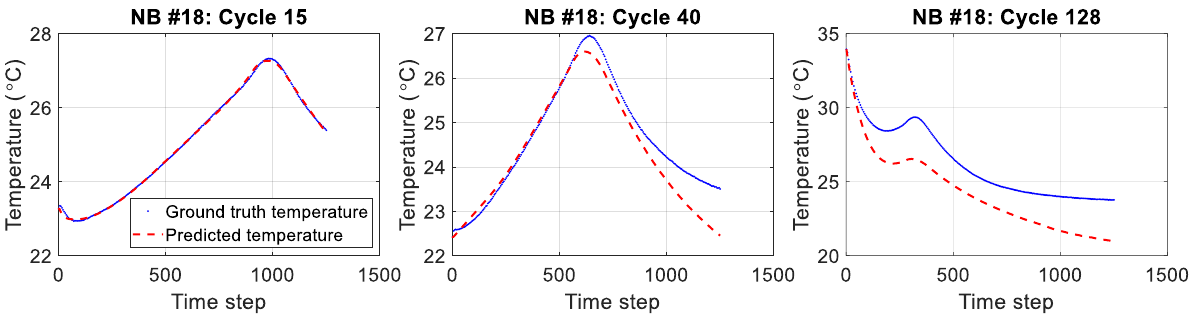}\\
\caption{Prediction performance of the developed ECTM with NB \#18 data (ambient temperature 24 $^\circ$C). Left: The fitting performance of ECTM on the base Cycle 15; Middle and Right: Predicted temperatures from identified ECTM for Cycle 40 (a medium degradation with 5.62\% capacity fade) and Cycle 128 (a major degradation  with 24.16\% capacity fade).} \label{fig:NB_18}
\end{figure*}
\begin{figure}[tbh]
\centering
\includegraphics[width=0.8\columnwidth]{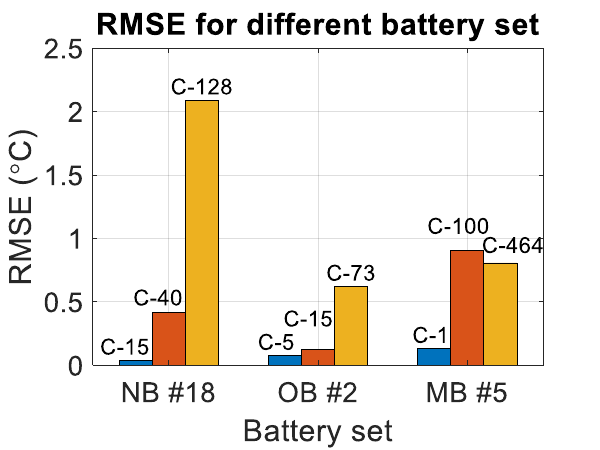}\\
\caption{The RMSE between predicted temperature profiles from ECTM and the ground-truth profiles for different batteries, where C-$k$ above the bars represents the Cycle $k$.
} \label{fig:rmse}
\end{figure}
\begin{figure*}[tbh]
\centering
\includegraphics[width=1.9\columnwidth]{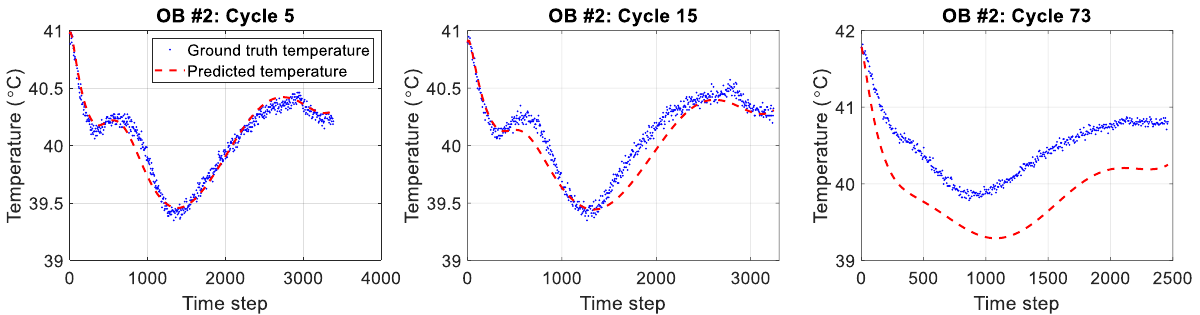}\\
\caption{Prediction performance of the developed ECTM with OB \#2 data (ambient temperature 40 $^\circ$C). Left: The fitting performance of ECTM on the base Cycle 5; Middle and Right: Predicted temperatures from identified ECTM for Cycle 15 (a medium degradation with 4.30\% capacity fade) and Cycle 73 (a major degradation  with 27.40\% capacity fade).
} \label{fig:ob_2}
\end{figure*}

\section{Parameter identification}
\label{sec:parameter_identification}
This section presents the least-squares method to identify the model parameters $\Theta$ as given in Section \ref{subsec:thermal_model}. The optimization problem is formulated as minimizing the sum of the squared errors between the observed temperature values and the model's predictions (cf. \eqref{eq:linear_compact}):
\begin{equation}
\min_{\Theta}\sum_{k=1}^{K}\left(T^o_s[k]-\sum_{j=1}^{m}\theta_jx_{j}[k-1]\right)^2,
\end{equation}
where $K$ is the total number of experiment data samples gathered from \textit{one charging or discharging cycle}, and $T_s^o[k]$ is the true surface temperature observed at the $k$-th timestep.
The above equation can also be compactly written into
\begin{equation}
\min_\Theta  \quad \lVert A\Theta-\bar{T}^o_s \rVert^2, \label{eq:matrix}
\end{equation}
where the coefficient matrix $A$, model parameter $\Theta$, and true temperature $\bar{T}^o_s$ in the above equation are defined as:
\begin{align}
A \coloneqq \left[\begin{matrix}
x_{1}[1] & x_{2}[1] \cdot \cdot \cdot & x_{m}[1]\\
x_{1}[2] & x_{2}[2] \cdot \cdot \cdot & x_{m}[2]\\
\cdot & \cdot \cdot \cdot \cdot \cdot & \cdot\\
\cdot & \cdot \cdot \cdot \cdot \cdot & \cdot\\
x_{1}[K] & x_{2}[K] \cdot \cdot \cdot & x_{m}[K]\\
\end{matrix}\right] \in \mathcal{R}^{K \times m}; \nonumber
\end{align}
\begin{align}
&\Theta^\top \coloneqq \left[\begin{matrix}
\theta_{1} &
\theta_{2} &
\cdots &
\theta_{m}
\end{matrix}\right]\in \mathcal{R}^{1 \times m}; \nonumber \\ & 
(\bar{T}_s^o)^\top \coloneqq \left[\begin{matrix}
\bar{T}_s^o[1] &
\bar{T}_s^o[2] &
\cdots &
\bar{T}_s^o[K]
\end{matrix}\right]\in \mathcal{R}^{1 \times K}. \nonumber
\end{align}
The unconstrained optimization in \eqref{eq:matrix} is a standard least-squares problem with closed-form solutions:
\begin{equation}
\Theta^{*}=\left(A^\top A\right)^{-1}A^\top \bar{T}_s^o.
\end{equation}

When prior knowledge on the feasible sets of $\Theta$ is available, it can be easily incorporated into \eqref{eq:matrix} to render it a constrained least-squares problem:
\begin{equation}
\min_{\Theta\in \Omega } \quad \lVert A\Theta-\bar{T}^o_s \rVert^2, \label{eq:matrix_constrained}
\end{equation}
where $\Omega\subseteq \mathcal{R}^{m}$ is the feasible set for the parameter. For such standard constrained least-squares problems, extensive off-the-shelf solvers are available, such as the trust-region reflective (TRR) solver \cite{ahsan2017system}. Moreover, these solvers are computationally efficient to identify the parameters of the proposed ECTM with one-shot cycling data and are fast enough for real-time applications.

\section{Results and discussion}
\label{sec:results}
\begin{figure}[tbh]
	\centering
	\includegraphics[width=0.9\columnwidth]{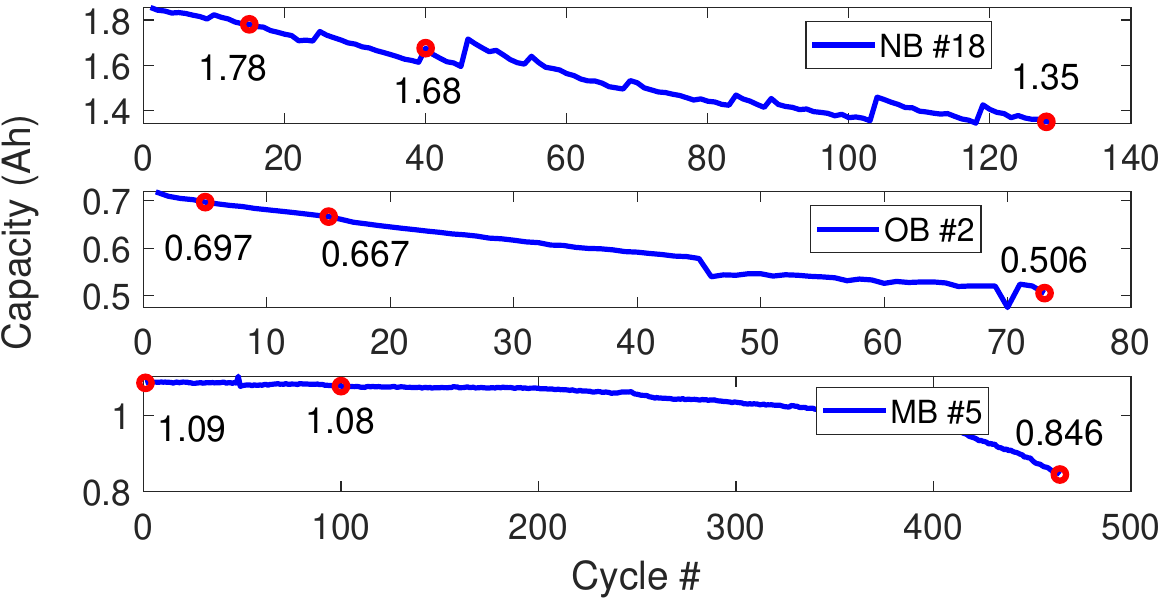}\\
	\caption{Capacity values of selected battery cells over cycles (red dots show the selected cycle numbers for study). 
	} \label{fig:capacity_fade}
\end{figure}

In this section, we test the validity of the proposed ECTM with batteries from three distinct benchmark datasets: (a) NASA aging dataset (NB) \cite{saha2017battery}; (b) Oxford battery degradation dataset (OB) \cite{birkl2017oxford}; and (c) MIT datasets (MB) \cite{severson2019data}. We have selected the charging cycle data collected from NB \#18, OB \#2, and MB \#5 for study. The NB \#18 was obtained from experiments performed on 18650 cells with 2 Ah capacity at 24 $^\circ$C. Constant-current constant-voltage (CCCV) protocol was applied for battery cycles until the battery reached a capacity fade of 30\%. For the OB, 740 mAh pouch cells were cycled with a CCCV protocol until reaching a 30\%  capacity fade in a temperature-controlled chamber (40 $^\circ$C). Finally, for MB \#5, commercial LFP/graphite cells were cycled with fast charging conditions until 20\% capacity fade was reached at an ambient temperature of 30 $^\circ$C. Fig. \ref{fig:capacity_fade} shows the capacity values of these three batteries over cycles, where red points indicate the cycles (one early-stage, one medium-stage, and one late-stage) selected for this study.

For identifying the model parameters $\Theta$, we have randomly selected Cycle 15, Cycle 5, and Cycle 1 from NB \#18, OB \#2, and MB \#5, respectively, as the base cycles. In light of the linearity of the ECTM in \eqref{eq:linear_model}, all the initial guesses for the solver are selected as 1. The identified parameters for the models of three batteries are given in Table \ref{table:parameters}.

\begin{table}[tbh]
\centering
\caption{Identified ECTM parameters with data of Cycle 15, Cycle 5, and Cycle 1, of NB \#18, OB \#2, and MB \#5.}
\label{table:parameters}
\begin{tabular}{lccccc}
\hline
Battery & \multirow{2}{*}{$\theta_1$} & \multirow{2}{*}{$\theta_2$} & \multirow{2}{*}{$\theta_3$} &\multirow{2}{*}{$\theta_4$} &\multirow{2}{*}{$\theta_5$} \\
Name & & & & &   \\
\hline
NB \#18 & 0.16 & 0.04 &  -0.25 & 0.25 & 0.22 \\
OB \#2 & 3.2 $\times 10^5$ & 2& -118.3 & 3.1 $\times 10^5$ & -3.6\ $\times 10^4$ \\
MB \#5 & 0.01 & -3.82 & 252.96 & -117.1 & 1.75 \\

\hline\hline
Battery & \multirow{2}{*}{$\theta_6$} &\multirow{2}{*}{$\theta_7$} &\multirow{2}{*}{$\theta_8$} & \multirow{2}{*}{$\theta_9$} \\
Name &  & & &  & \\

\hline
NB \#18  &-0.36 & 0.04 & 0.99 & 0.01\\
OB \#2 & 1.5 $\times 10^5$ & -0.02 & 0.99 &-1.2 $\times 10^5$\\
MB \#5 & -185.5 & 0.01 &0.99 &0.02\\

\hline
\end{tabular}%
\end{table}
\begin{figure*}[tbh]
\centering
\includegraphics[width=1.9\columnwidth]{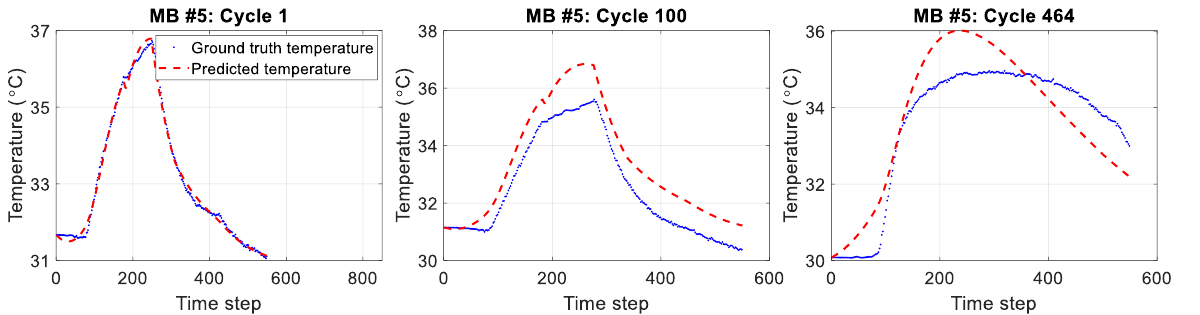}\\
\caption{Prediction performance of the developed ECTM with MB \#5 data (ambient temperature 30 $^\circ$C). Left: The fitting performance of ECTM on the base Cycle 1; Middle and Right: Predicted temperatures from identified ECTM for Cycle 100 (medium degradation with 0.92\% capacity fade) and Cycle 464 (major degradation with 22.38\% capacity fade).} \label{fig:MB_5}
\end{figure*}

\begin{figure*}[tbh]
\centering
\includegraphics[width=2\columnwidth]{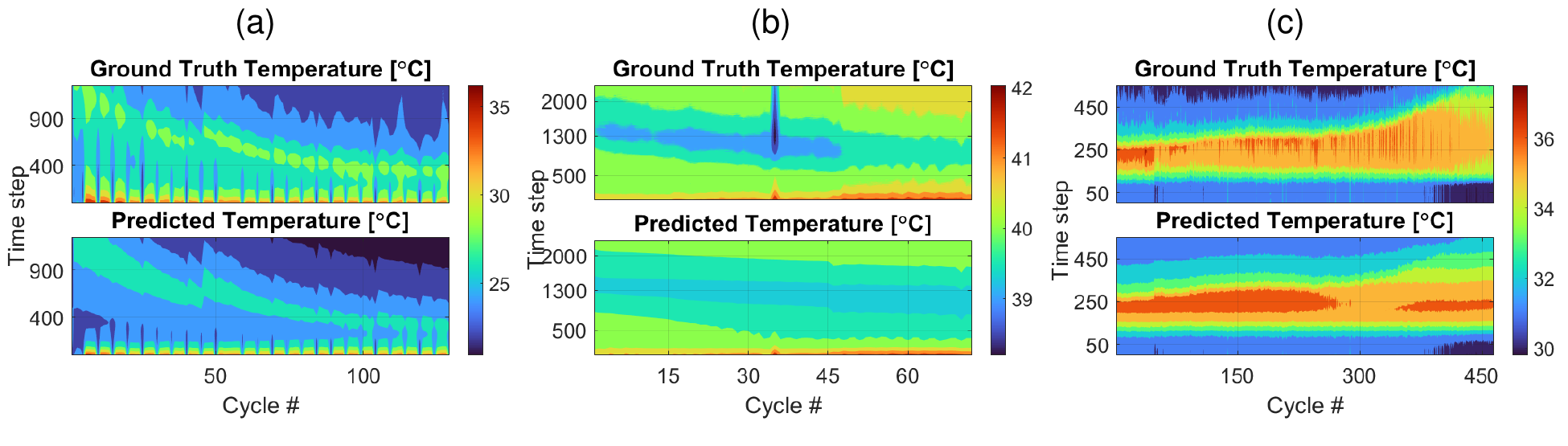}\\
\caption{The profiles across cycles for ground-truth temperatures (top) and predicted temperatures from ECTM identified from the base cycles (bottom) for NB \#18, (b) OB \#2, and (c) MB \#5.
} \label{fig:contour}
\end{figure*}

The left plot in Fig. \ref{fig:NB_18} shows the training performance of the ECTM for the temperature profile in base Cycle 15 of NB \#18. We observe that the estimated parameters (given in Table \ref{table:parameters}) for NB \#18 can give a highly accurate fit to the data. Afterwards, we use the model identified from base Cycle 15 to predict the temperature profiles for Cycle 40 (short-term) and Cycle 128 (long-term), given in the middle and right plots of Fig. \ref{fig:NB_18}, respectively. To assess the accuracy of the model, we compute the root mean square error (RMSE) between the predicted and ground-truth temperatures, as shown in the left group of bars in Fig. \ref{fig:rmse}. Overall, the trained ECTM shows satisfactory generalization performance into future cycles with high prediction accuracy, especially considering that in Cycle 128, the battery has degraded dramatically with a capacity fade of 24.16\%.  

We conduct similar experiments for OB \#2 and MB \#5, where the ambient temperatures are 40 $^\circ$C and 30 $^\circ$C, respectively. We intend to assess the performance of the proposed ECTM under such elevated ambient temperatures. Fig. \ref{fig:ob_2} shows the test results for OB \# 2. With the parameters estimated based on the data from base Cycle 5 (left plot), the predictions from the developed ECTM for Cycle 15 (medium degradation) and Cycle 73 (major degradation) show great accuracy, as in the middle and right plots. Similarly, for battery MB \#5, the ECTM can also generalize well into future cycles that are far from the training cycle, as shown in Fig. \ref{fig:MB_5}. Note that the total cycle number of each battery is different due to their difference in types and experiment conditions: OB \#2 has a total cycle number less than 80 due to the harsh ambient conditions, whereas MB \#5 sustains significantly more testing cycles. Fig. \ref{fig:contour} compares the ground truth and predicted temperature profiles from the ECTM identified using base cycle data for each battery. It clearly shows that the predicted ones are highly aligned with the ground-truth ones. This indicates the robustness of the ECTM across different battery types and operating conditions, even under severe degradation scenarios.

\section{Conclusions}
This work proposed a novel equivalent-circuit electro-thermal scheme to model battery temperatures. A major contribution of this method lies in the approximation of the complex heat generation inside the battery with a physics-based data-driven nonlinear (polynomial) function of SOC, current, and terminal voltage. Moreover, the presented thermal model is formulated as a linear form in which the parameter estimation is computationally efficient. The efficacy of the model is tested using the benchmark NASA, Oxford, and MIT datasets, where the ambient temperatures vary significantly. Simulation results show that the developed battery temperature model can predict the temperature profiles accurately under various degradation conditions and ambient temperatures. These observations validate the effectiveness and robustness of the proposed equivalent-circuit method in modeling battery temperatures. 

\section{Acknowledgment}
M. A. Chowdhury acknowledges the support of Distinguished
Graduate Student Assistantships (DGSA) from Texas Tech University. Q. Lu acknowledges the new faculty startup funds from Texas Tech University. The authors acknowledge the support from the National Science Foundation under Grant No. 2340194.

\bibliographystyle{ieeetr}
\bibliography{thermal_model_ref}
\end{document}